\documentclass{article}
\usepackage[margin=1in]{geometry} 
\usepackage{graphicx} 
\usepackage[numbers]{natbib}
\usepackage{amsmath}
\usepackage{xcolor}
\usepackage{booktabs} 
\usepackage{authblk} 
\usepackage{placeins}
\usepackage{pdfpages}

\newcommand{\com}[1]{\textcolor{black}{#1}}

\newcommand{\tun}{T\textsubscript{1}} 
\newcommand{\tdeux}{T\textsubscript{2}}
\newcommand{\tdeuxetoile}{T\textsubscript{2}*}
\newcommand{\bun}{B\textsubscript{1}}
\newcommand{\bzero}{B\textsubscript{0}}
\newcommand{\mzero}{M\textsubscript{0}}

\begin{document}

\begin{center}
    \Huge \textbf{Submitted to Magnetic Resonance in Medicine}
\end{center}

\vspace{1em}

\title{\textbf{Relaxometry and contrast-free cerebral microvascular quantification using balanced Steady-State Free Precession MR Fingerprinting\protect}}

\date{August 2024}

\author[1]{Thomas Coudert}
\author[2]{Aurélien Delphin}
\author[1]{Antoine Barrier}
\author[1,3]{Loïc Legris}
\author[1]{Jan M. Warnking}
\author[2]{Laurent Lamalle}
\author[4]{Mariya Doneva}
\author[1]{Benjamin Lemasson}
\author[1]{Emmanuel L. Barbier}
\author[1]{Thomas Christen}

\affil[1]{Univ. Grenoble Alpes, INSERM, U1216, Grenoble Institute Neurosciences, GIN, Grenoble, France}
\affil[2]{Univ. Grenoble Alpes, INSERM, US17, CNRS, UAR3552, CHU Grenoble Alpes, IRMaGe, Grenoble, France}
\affil[3]{Univ. Grenoble Alpes, Stroke Unit, Department of Neurology, CHU Grenoble Alpes, Grenoble, France}
\affil[4]{Philips Innovative Technologies, Hamburg, Germany}

\maketitle

\begin{abstract}
This study proposes a novel, contrast-free Magnetic Resonance Fingerprinting (MRF) method using balanced Steady-State Free Precession (bSSFP) sequences for the quantification of cerebral blood volume (CBV), vessel radius (R), and relaxometry parameters (\tun, \tdeux, \tdeuxetoile) in the brain. The technique leverages the sensitivity of bSSFP sequences to intra-voxel frequency distributions in both transient and steady-state regimes. A dictionary-matching process is employed, using simulations of realistic mouse microvascular networks to generate the MRF dictionary. The method is validated through in silico studies and in vivo experiments on six healthy subjects, comparing results with standard MRF methods and literature values. The proposed method shows strong correlation and agreement with standard MRF methods for \tun{} and \tdeux{} values. High-resolution maps provide detailed visualizations of CBV and microvascular structures, highlighting differences in white matter (WM) and grey matter (GM) regions. The measured GM/WM ratio for CBV is 1.91, consistent with literature values. This contrast-free bSSFP-based MRF method offers an innovative approach for quantifying CBV, vessel radius, and relaxometry parameters. Further validation against DSC imaging and clinical studies in pathological conditions is warranted to confirm its clinical utility.
\end{abstract}

\section{Introduction}\label{intro}

Estimations of brain microvascular properties (blood volume, blood flow, microvessel diameter, capillary transit time, etc.) are crucial for diagnosing, staging, and evaluating therapies in various pathologies like stroke and tumors \cite{HGG_mri,braintumor_DSC,stroke_perfusion}. 

A common MR method to assess microvascular properties is the Dynamic Susceptibility Contrast (DSC) approach which relies on MR signal variations following an intravenous injection of gadolinium (Gd) based contrast agents (CA). The CA induces magnetic susceptibility differences between the blood compartment and the surrounding tissues. This leads to broadening the distributions of MR frequencies inside the imaging voxels, and a corresponding reduction of gradient echo (GRE) signal intensities proportional to the voxel blood volume in the absence of large vessels\cite{Tropres2015}. Simple biological models can be used to retrieve cerebral blood flow (CBF) or mean arterial transit time (MTT) and more advanced models can estimate capillary transit times and maximum oxygen extraction fraction (OEF)\cite{JespersenOEF}. Interestingly, signal magnitudes from spin echo (SE) sequences are also reduced due to water diffusion effects. Combining GRE and SE signal variations can provide information sensitive to microvessel density and size.\cite{Stadlbauer2017VascularHL, emblem_vessel} Absolute quantification of vascular parameters using DSC is however difficult \cite{CalamanteDSC}. It requires estimations of local Arterial Input Functions (AIF) and signal deconvolutions. The main drawback of DSC is the use of bolus of CA which not only limits spatial resolution due to the need for dynamic measurements but also increases the risk of nephrogenic systemic fibrosis and potential gadolinium retention in brain or kidney tissue\cite{iyad2023gadolinium}.

Several methods based on the Blood Oxygen Level Dependent (BOLD) effect have been proposed to avoid CA injection.\cite{christen_blood_oxy2012,BIONDETTI2023120189}. Many studies have used \tdeuxetoile{} maps as surrogates for blood volume and/or oxygenation estimates. These maps are computed by considering Lorentzian frequency distributions inside the voxels and corresponding mono-exponential signal decays from multi-echoes GRE sequences. These baseline maps are however not fully reliable as many non-vascular effects (\tdeux, shim) also contribute to the signal decay which often deviates from the simple exponential model. A more advanced analysis of the baseline BOLD effect, named quantitative BOLD has also been proposed \cite{qbold,christen_tissueoxy_mri,christen_qbold}. It relies on modeling the frequency distributions using statistical models of blood vessels represented as ensembles of isotropically oriented straight magnetic cylinders. Accounting for other signal contributions, small deviations from monoexponential MR signal decay at short echo times allow the separation of cerebral blood volume (CBV) and blood oxygen saturation (SO\textsubscript{2}) contributions. This is however possible only at very high SNR and the cylinder model might not be able to represent all types of vascular structures.\cite{Sedlacik2010} Recently, interesting results have been reported on non-contrast blood volume and blood arrival estimates using either internal resting-state fluctuations\cite{christen_moyamoya} or externally controlled gas challenges to mimic boluses of paramagnetic blood. However, these approaches also suffer from low SNR and thus low spatial resolution. 

Most of the previous MR studies designed for microvascular estimates have been based on GRE sequences. It has been known for decades that other types of sequences are more sensitive to frequency shifts. In particular, the balanced steady-state free precession (bSSFP) sequences have a signal response (magnitude and phase) that highly depends on the frequency offset $\delta \! f$\cite{Scheffler2003PrinciplesAA, miller_fmri_bssfp}. In comparison, the frequency response functions of GRE are always flat in magnitude and linear in phase. When measuring with a bSSFP sequence the signal from a voxel that contains a large distribution of resonance frequencies, the variations often deviate from exponential decay and can in some cases produce spin echo-like behaviors \cite{schefflerTrueFISP,leupold2018}. Moreover, the bSSFP response functions are known to depend on the sequence parameters and to differ between transient and steady states. It is thus possible to induce changes in signal evolution by changing the flip angle (FA) or repetition time (TR) rather than trying to modify the in vivo distributions of frequencies using CA injection. The high sensitivity of bSSFP to microvascular properties has already been observed in several BOLD fMRI experiments\cite{miller_fmri_bssfp}. However, it is also clear that quantitative estimates from bSSFP-type sequences are difficult to obtain as the signals also depend on various non-vascular parameters including \tun, \tdeux, or \bun.

The Magnetic Resonance Fingerprinting (MRF\cite{Ma2013}) framework was proposed 10 years ago to allow the extraction of multiple parameters simultaneously from complex MR sequences in their transient states. Matched to a dictionary of signals obtained in silico, even highly undersampled images can produce reliable quantitative relaxometry maps.  bSSFP type sequences have been analyzed in the first MRF study to produce \tun, \tdeux, \mzero, and $\delta \! f$ maps. However, FISP (Fast Imaging with Steady Precession) GRE sequences have since been preferred because of lighter dictionary generation and analysis. Two recent studies\cite{Wang2019,boyacioglu_quadratic_2021} have proposed to analyze bSSFP sequences for additional \tdeuxetoile{} estimates using Lorentzian intra-voxel frequency distributions in the dictionaries. In parallel, the MRF framework has also been used to assess vascular properties. An ASL-based sequence combined with MRF (MRF-ASL\cite{su_mrf_asl}) has shown promising results on CBF and bolus arrival time (BAT) estimations validated against DSC MRI. Additionally, multi-echo spin- and gradient-echo sequences acquired pre and post-CA injection allow the estimation of CBV, mean vessel radius (R), and SO\textsubscript{2} within the MRF framework (MR vascular Fingerprinting or MRvF \cite{Christen2014}). Compared to standard relaxometry MRF, these sequences provide only a few fully sampled images but the dictionaries involve realistic biophysical representations of the vascular networks. The magnetic field perturbations due to the magnetic susceptibility distributions and the phase accumulation due to water diffusion are also computed. In that case, high-resolution maps of CBV, R, and SO\textsubscript{2} have been obtained in rats and humans \cite{Christen2014, Lemasson2016}.

The current paper hypothesizes that bSSFP sequences, used in an extended MRvF framework, can provide quantitative maps of \tun, \tdeux, \tdeuxetoile, M\textsubscript{0}, $\delta \! f$, CBV, and micro-vascular properties without CA injection. We studied the sensitivities of GRE and bSSFP sequences in silico and proposed an MRvF-bSSFP candidate for vascular estimates. We compared the results obtained in human volunteers using large multidimensional dictionaries generated from Lorentzian distributions of magnetic field or frequency distributions based on 3D vascular voxels, either artificially generated (standard approach using cylinders) or segmented from microscopy datasets (recently proposed in Delphin et al.\cite{Delphin2024}). 


\section{Methods}\label{matmet}
\subsection{Two-step MRF Dictionary generation}\label{dicogen}

Standard MRF dictionaries describing signal evolutions of unbalanced GRE sequences usually contain 3 dimensions (\tun, \tdeux, \bun)\cite{jiang2015mr}. To represent the resonance frequency shift sensitivity, dictionaries that describe bSSFP sequences need to take into account an extra dimension that describes the intra-voxel average frequency offset ($\delta \! f$). 
We first generated such a base dictionary $Dico_{Base}$(\tun, \tdeux, \bun, and $\delta \! f$) using in-house Python+Matlab code derived from a reference Bloch simulator \cite{Bloch_Simulator} for standard relaxometry sequences.  Simulations were performed at 3.0T with 20 \tun{} values (0.2 to 3.5s), 20 \tdeux{} values (10 to 600ms), 10 \bun{} values (0.7 to 1.2) and frequency offset $\delta \! f$ values (from -50 to 49 Hz with an increment of 1 Hz), keeping only signals for which \tun$>$\tdeux, resulting in a $390,000$ entries dictionary.

\begin{figure*}
\centerline{\includegraphics[  width=\textwidth  ]{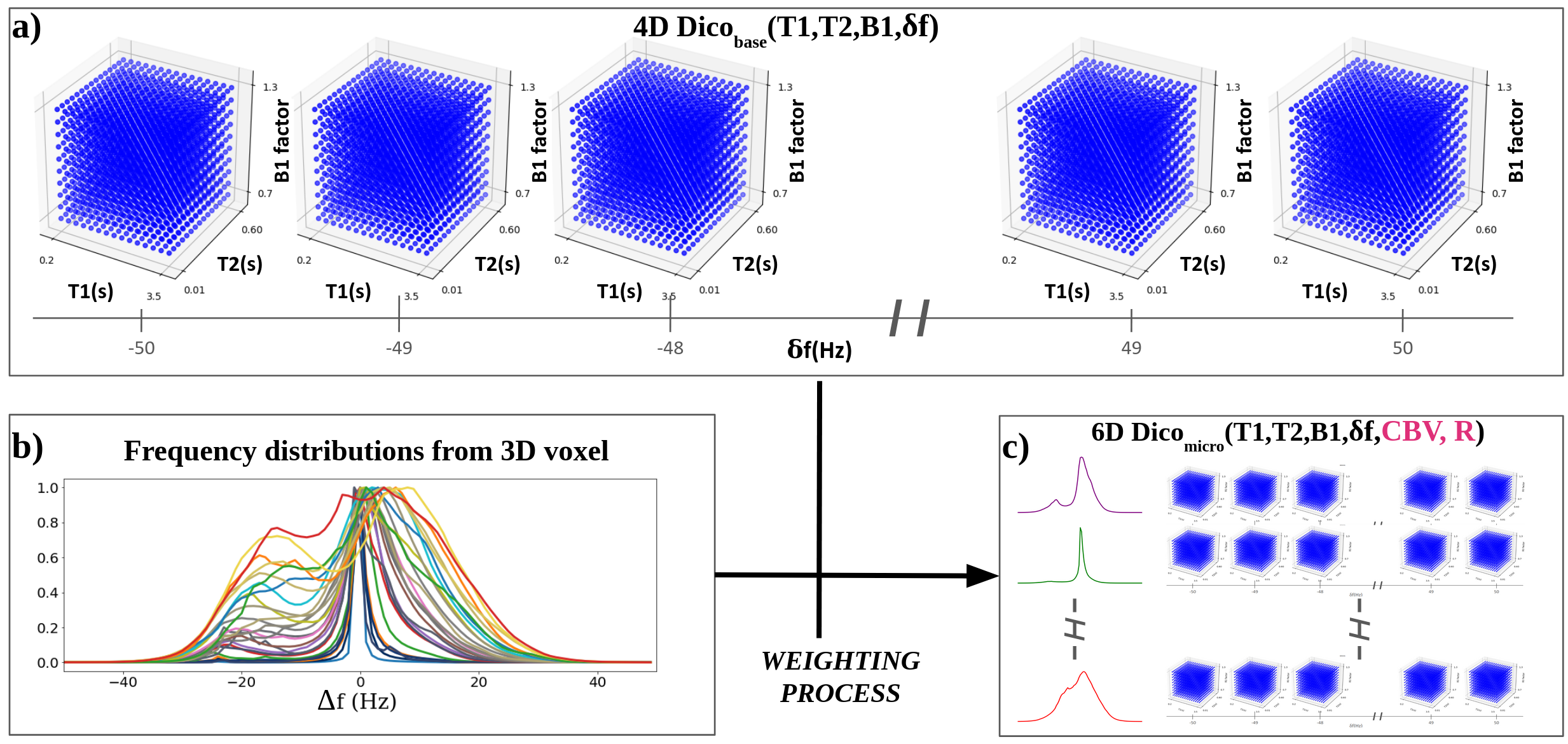}}
\caption{Two-step MRF simulations process. a)Simulations of a standard 4D Bloch dictionary with (\tun,\tdeux,\bun,$\delta \! f$) parameters grid. b)Use of frequency distributions from 3D voxels to weight the 4D dictionary resulting in a 6D dictionary (here shown for example at df=0Hz) (c) with CBV and R dimensions related to the 3D frequency distributions.\label{fig1}}
\end{figure*}

New dimensions need to be added to the dictionary when accounting for \tdeuxetoile{} or microvascular effects. In the first approximation, the influence of blood vasculature on signal evolutions can be seen through the changes in the voxel frequency shift distributions caused by magnetic sources\cite{miller_modeling_2008}. While it is possible to compute the corresponding voxel spatial magnetic field distributions and to sum the MR signals to come from multiple spatial locations\cite{Christen2014}, a faster solution needs to be proposed when the dictionaries also contain multiple relaxometry parameters and thus hundreds of millions of simulated signals. If the influence of water diffusion is low (\com{Figure S1 (supp)}), the voxel spatial frequency distributions can be simplified and it is possible to focus on the shapes of the frequency histograms only. Following the work of Wang et al\cite{Wang2019}, a \tdeuxetoile{}-like behavior, sensitive to microvascular properties, may simply be obtained by recombining several complex signals from the base dictionary according to the histogram of intra-voxel frequency values. Our $Dico_{Base}$ was simply weighted by pre-computed intra-voxel frequency distributions in a range of $\Delta \! f$ frequencies centered on $\delta \! f$ values (see Figure \ref{fig1}). We evaluated four methods: standard Lorentzian distributions and 3 approaches based on the simulation of an intravoxel microvascular network, with increasing biophysical complexity, as detailed below. These distributions were defined by specific parameters $\theta$ that can then be estimated in the MRF framework by matching into a new dictionary which depends on $(T_1, T_2, B_1, \delta \! f, \theta)$ (see \com{Figure S2 (supp)}).

\subsubsection{Lorentzian frequency distributions for \tdeuxetoile{} estimates (Dico\textsubscript{\tdeuxetoile})}
It is common to consider that the intra-voxel frequency distribution has a Lorentzian shape. This shape was chosen such that a monoexponentially decaying transverse signal evolution would be produced from a GRE sequence. This approximation defines the \tdeuxetoile{} relaxation time which corresponds to the width of the distribution (or the time constant of the exponential signal decay of unbalanced multi-gradient echo sequences). Here, we followed the work by Wang et al\cite{Wang2019}, by modeling the intra-voxel frequency distribution with probability density function given by:
\begin{equation}\label{eq:1}
	L(\Delta \! f, \Gamma, \delta \! f) = \frac{\frac{\Gamma}{2}}{(\Delta \! f - \delta \! f)^2 + (\frac{\Gamma}{2})^2}
\end{equation}
where $\Delta \! f$ is the frequency-dependent signal values of the $Dico_{Base}$, $\delta \! f$ the mean frequency of the voxel and $\Gamma$ the full width at half maximum of the distribution (ranging from 0 to 20 Hz with an increment of 1 Hz). 
Using the two-step dictionary generation process described above, an expanded dictionary that depends on $(T_1,T_2,B_1,\delta \! f,\Gamma)$ is obtained by weighting each complex signal contribution by the spin density at the corresponding frequency.

This dictionary can be then interpreted as Dico\textsubscript{\tdeuxetoile}$(T_1,T_2,B_1,\delta \! f,T_2^*)$ by using:

\begin{equation}\label{eq:2}
	\frac{1}{T_2^*} = \frac{1}{T_2} + \pi\Gamma
\end{equation}

To ensure fully defined distributions in the convolution process only values in the range $-30$ Hz $<\delta \! f<30$ Hz were used in the expanded dictionary, leading to a total number of 4,680,000 entries.

\subsubsection{Realistic frequency distributions for vascular estimates}

The in vivo intra-voxel magnetic field distribution is usually more complex than a Lorentzian distribution. It is possible to compute the magnetic field distributions produced by 3D vessel structures using a fast Fourier-based approach\cite{salomir_fast_2003,
marques_application_2005}. The corresponding $\delta \! f$ distributions thus depend on vascular characteristics such as CBV, R, SO\textsubscript{2}. The simulations of the distributions were made with a main magnetic field of 3.0T oriented in the Z direction, with the SO\textsubscript{2} in each voxel set at 70\% and the micro-vascular hematocrit fraction at 0.85$\times$42\%\cite{hematocrit}. 

\begin{figure*}
\centerline{\includegraphics[  width=\textwidth  ]{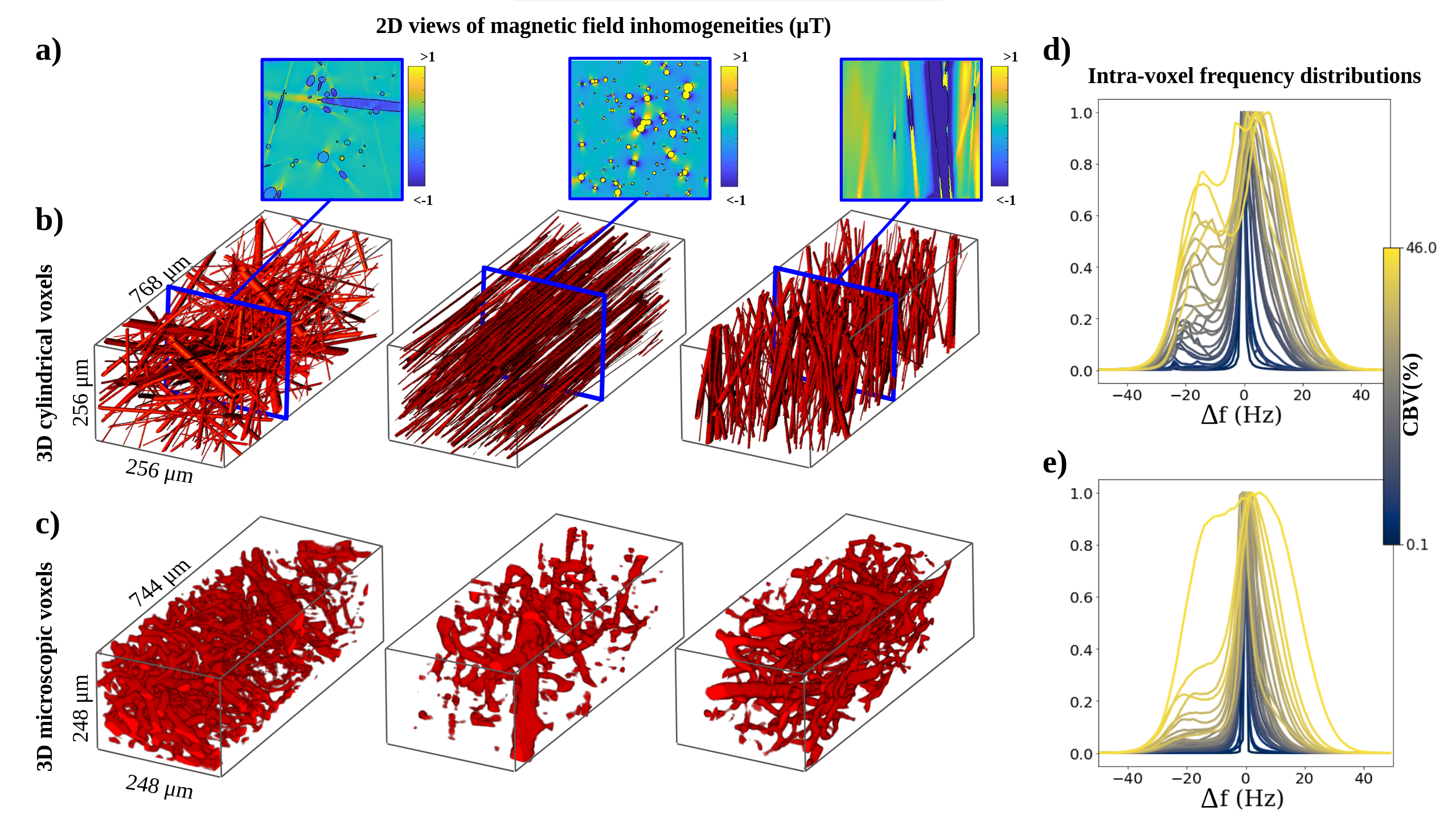}}
\caption{Magnetic field simulations (a) in vascular voxels with fixed blood susceptibility value, leading to a set of intra-voxel frequency distributions: using synthetic 3D-cylindrical vascular voxels (b, d) and using microscopy-segmented 3D vascular voxels (c, e). For the distributions, color code corresponds to (CBV,R) and is mainly influenced by changes in the CBV which varies between 0.1 and 46\%.\label{fig3}}
\end{figure*}

\paragraph{3D cylindrical voxels ($Dico_{Cyl}$)}
Straight cylinders are common approximations for blood vessel shapes in MR simulations. We thus generated 3D voxels containing multiple straight cylinders with variable radii, using Matlab. 2500 combinations of (CBV, R) were chosen, with manually distributed ranges for each parameter, where the mean radius R was set as the center of a Gaussian distribution of vessel radii in a given voxel, resulting in 1878 3D voxels taking into account that the geometrical constraints of the cylinder generation can not always accommodate all (CBV, R) combinations. Vessels were isotropically oriented in 60\% of the voxels and anisotropically oriented along B0 and in the transverse plane in 20\% each. After validating that the voxel size does not significantly influence the distribution of the intra-voxel magnetic inhomogeneities as long as the size ratios between the axes and their orientation were respected (\com{Figure S3 (supp)}), the voxel size was fixed at 256x256x768µm\textsuperscript{3}  with a 2.0x2.0x2.0µm\textsuperscript{3} resolution for vascular structures simulations. From these 3D voxels, magnetic field distributions were computed using a Fourier transform (see Figure \ref{fig3}). The corresponding 1878 frequency distributions were then used to expand the 390,000 entries $Dico_{Base}$ dictionary, leading to a 732,420,000 entries vascular dictionary $Dico_{Cyl}(T_1,T_2,B_1,\delta \! f,CBV,R)$. Only $-30$ Hz $<\delta \! f<30$ Hz values were used for matching in the expanded dictionary, with simulations computed on the fly during the matching process for efficiency.

\paragraph{3D microscopic voxels ($Dico_{Micro}$)}

As described previously\cite{Delphin2023}, it is possible to use microscopy-derived vascular networks as input for the magnetic field simulations. Here we use 2500 3D voxels segmented from multiple open-access datasets of whole brain, healthy mouse vascular networks as a basis to create a dictionary of signals with 3D-resolved MR simulations (see Figure \ref{fig3}). The voxel size is 248x248x744µm\textsuperscript{3} with a 2.0x2.0x2.0µm\textsuperscript{3} resolution. The voxels were chosen among a dataset of 30,000 segmented voxels such that the (CBV, R) values distribution was as close as possible to the 3D cylindrical voxels (CBV, R) distribution. At the end, the $Dico_{Micro}(T_1,T_2,B_1,\delta \! f,CBV,R)$ contains 975,000,000 entries and only $-30<\delta \! f<30$ Hz values were used for matching. 

\paragraph{Rectangular distributions for additional contributions}\label{contrib}
bSSFP sequences are very sensitive to \bzero{} variations and other sources of field inhomogeneities can influence the final vascular estimates. To consider this effect in our reconstructions, we also convolved each of the 2500 microscopic distributions with seven rectangular-shaped distributions with different widths representing spatial linear gradients across the voxel (Figure \ref{fig10}c-e). This finally leads to 17,500 distributions with a new dimension ("\bzero{} gradient") in the convoluted dictionary, determining the width of the additional top-hat function, and ranging between $10^{-5}$ and $10^{-2}$ T m\textsuperscript{-1}. Given the large size of the final dictionary (6,825,000,000 entries), the matching process was constrained here by a ground truth B1 map acquired on the same subject to reduce computation cost as it was introduced by Ma et al\cite{ma2017slice}.

\subsection{In silico study}
We used our dictionary simulations to compare in silico the \bzero{} inhomogeneity sensitivity of FISP and bSSFP sequences. Different sets of sequence parameters were used for FISP-type sequences to look at multi-echoes signal decay in response to frequency distributions obtained with and without contrast agent (CA) contribution(\com{Figure S4 (supp)}. In the latter case, the magnetic susceptibility of the blood compartment at equilibrium was set to $5.5$ ppm. The frequency distributions were computed from 27 3D vascular voxels with varying SO\textsubscript{2}, CBV, R values, at 3 Tesla, with a blood hematocrit fraction of $0.85 * 42\%$ and no water diffusion effect. To examine bSSFP-type sequences' sensitivity to non-contrast frequency distribution, varying sequence parameters were also used to simulate multi-echo signal decay, and both the transient-state and steady-state parts of the signal were studied. In every case, the magnitude and phase frequency response profiles were also calculated (at an echo time of TR/2).

\subsection{MR Data Acquisition}\label{acq}
In vivo acquisitions were realized on 6 healthy volunteers (28.0$\pm$5.5 years old, 3 males and 3 females) using a 32-channel head receiver array on a Philips 3T Achieva dStream MRI at the IRMaGe facility (MAP-IRMaGe protocol,  NCT05036629).  This study was approved by the local medical ethics committee and informed consent was obtained from all volunteers prior to image acquisition. The proposed MRvF-bSSFP sequence was based on an IR-bSSFP acquisition. 260 repetitions were acquired (TR=21 ms, TE=7 ms) with FA linearly increasing from 7$^\circ$ to 70$^\circ$\cite{Gomez} and a quadratic phase cycle of 10$^\circ$ increments with additional $(0,\pi)$ cycling between subsequent acquisitions.

For our proof of principle study, the acquisitions were performed using quadratic variable density spiral sampling (12 interleaves out of 13), matrix size=192x192x4, voxel size=1.04x1.04x3.00mm$^3$ for a total scan duration of 2 minutes per slice. For comparison, an MRF IR-spoil sequence\cite{Gomez} was used to acquire \tun, \tdeux, and \mzero{} quantitative maps. For \tdeuxetoile{} validation, a standard multi-echo GRE (MGRE, TR=70 ms, 10 echoes TE\textsubscript{1}=3.8 ms and $\Delta$TE = 6.4 ms) sequence was acquired. Finally, an anatomical T1w MRI (spin-echo sequence with FA=90° and TR=600ms, a matrix size of 512x512x327 with a slice thickness of 0.55mm and a spatial resolution of 0.447x0.447mm$^2$) was acquired and used for regions of interest (ROIs) delineation.

In one subject, the MRvF-bSSFP sequence was also acquired with cartesian sampling (compressed sense factor of 4) and high resolution in 2D to reduce shim and partial volume effects with a matrix size of 256x256x1 and a voxel size of 0.78x0.78x3.00mm$^3$, leading to a total scan time of 12 minutes. Additionally, a DREAM \cite{DREAM} sequence was applied to obtain a reference B1 map.

\subsection{MR Data Processing}\label{dataproc}
All processing was performed with Python. An MRF standard dictionary-based matching method was used. The dot product of each acquired fingerprint with the whole dictionary was computed and the entry yielding the highest value was kept as the best match. Due to constraints in computational power and storage size, the matching process was batched over the acquisition size and also over the number of signals of the dictionary in the case of convolved dictionaries by convolving distribution on the fly (see section \ref{dicogen}). Combining simulations and matching, for a 4-slices acquisition with the resolution detailed in section \ref{acq}, the reconstruction process takes about 2 hours on a single NVIDIA Quadro RTX 8000 with 48Go of memory using the JAX library\cite{jax2018github}. Concerning reference acquisitions, \tdeuxetoile{} values were derived from exponential fitting to the MGRE signal decay. \tun{} and \tdeux{} quantitative reference values were computed by a classical matching approach using a Bloch-simulated dictionary with the same (\tun, \tdeux, \bun) ranges as the $Dico_{Base}$ in section \ref{dicogen}. 

\subsection{Image analysis}
MRF \tun, \tdeux{} and \tdeuxetoile{} values for the grey matter (GM) and the white matter (WM) regions were calculated from automated ROIs generated using the Otsu’s thresholding method\cite{otsu} in Matlab using anatomical T1WI as reference. CBV and R values for the GM, and WM regions were computed using the same method, and values inside the superior sagittal sinus (SS) vein were also computed from manual ROI validated by a neuroscientist. The reference ROI values for relaxometry estimates were computed for the MR reference sequences described in section \ref{acq}. 
The accuracy and precision of the proposed method were assessed by comparing relaxometry \tun\&\tdeux{} quantitative estimates with acquired validation maps from IR-FISP MRF, using Bland-Altman analysis. Result maps were smoothed with a Gaussian filter (with a Gaussian kernel of standard-deviation $\sigma$=0.4 pixel) before computation only to improve plot clarity in Figure \ref{fig9}c,e.


\section{Results}
\subsection{In silico study}

\begin{figure*}
\centerline{\includegraphics[  width=\textwidth  ]{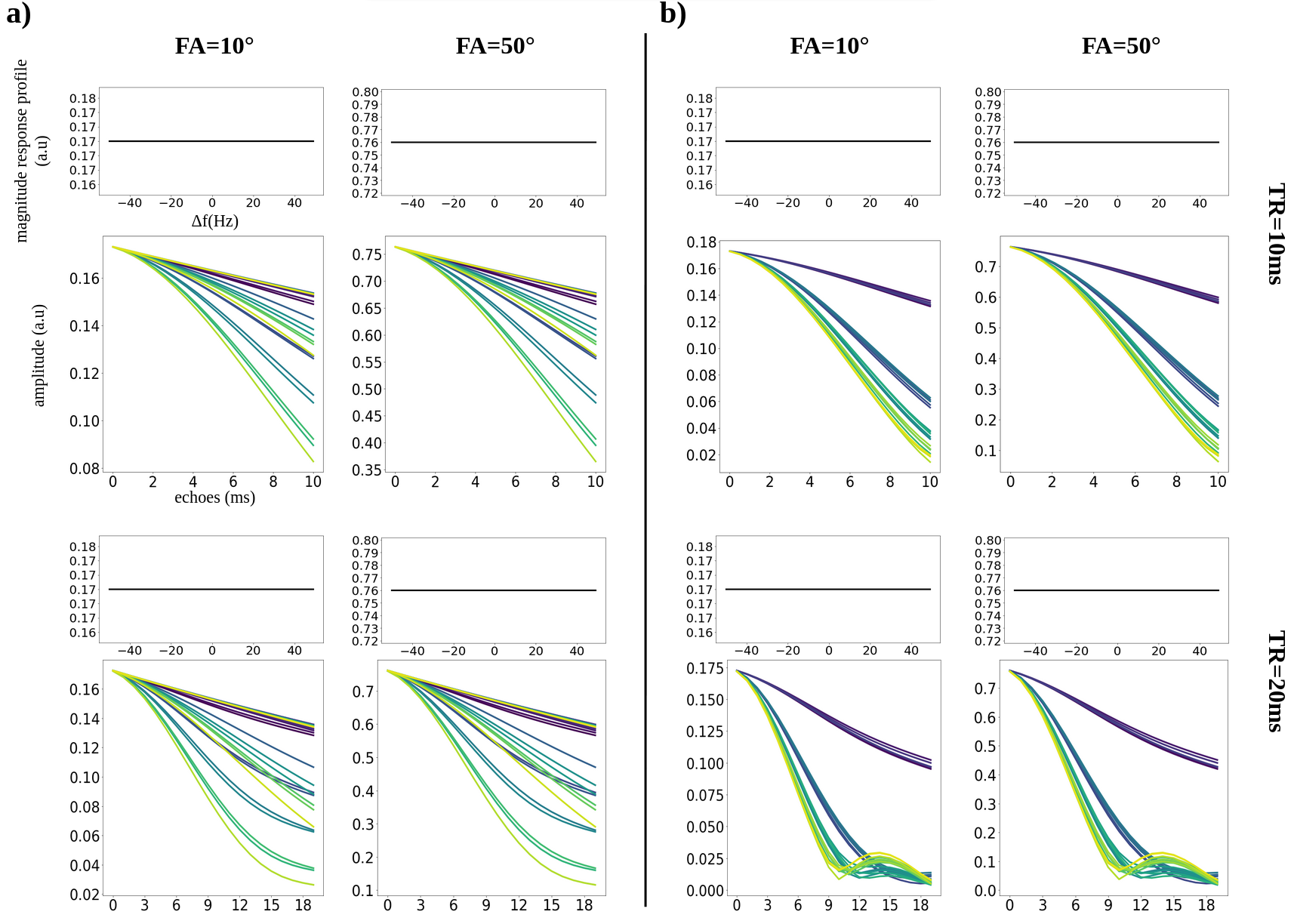}}
\caption{In silico study of the FISP signal response to different intra-voxel frequency distributions. The magnitude response profile and signal decays associated with intra-voxel frequency distributions are shown without CA in a) and with CA in b). The magnitude response profile is shown for TE=TR/2 for TR=10ms and TR=20ms, and average tissue properties of grey matter at 3T (GM, \tun=1300 ms, and \tdeux=80 ms) and $\delta \! f$=0Hz. The color code is primarily related to the CBV value.\label{fig4}}
\end{figure*}

In Figure \ref{fig4}a, we show the frequency distributions obtained in voxels with realistic microvascular networks with different CBV, R, and SO\textsubscript{2} properties. The distributions from the same voxels containing the contrast agent are represented in Figure \ref{fig4}b, highlighting peak broadening. Color code corresponds to (CBV,R,SO2) and is sorted by CBV values. As expected, the frequency magnitude response profile of the spoiled sequence is flat for every set of sequence parameters (TR, FA)leading to quasi-mono-exponential decays in multi-echo FISP sequences. With CA, signal changes at longer echo times reveal CBV differences. Note that for the CA injection case, distributions are highly truncated on the fixed range of frequency.

\begin{figure*}
\centerline{\includegraphics[  width=\textwidth  ]{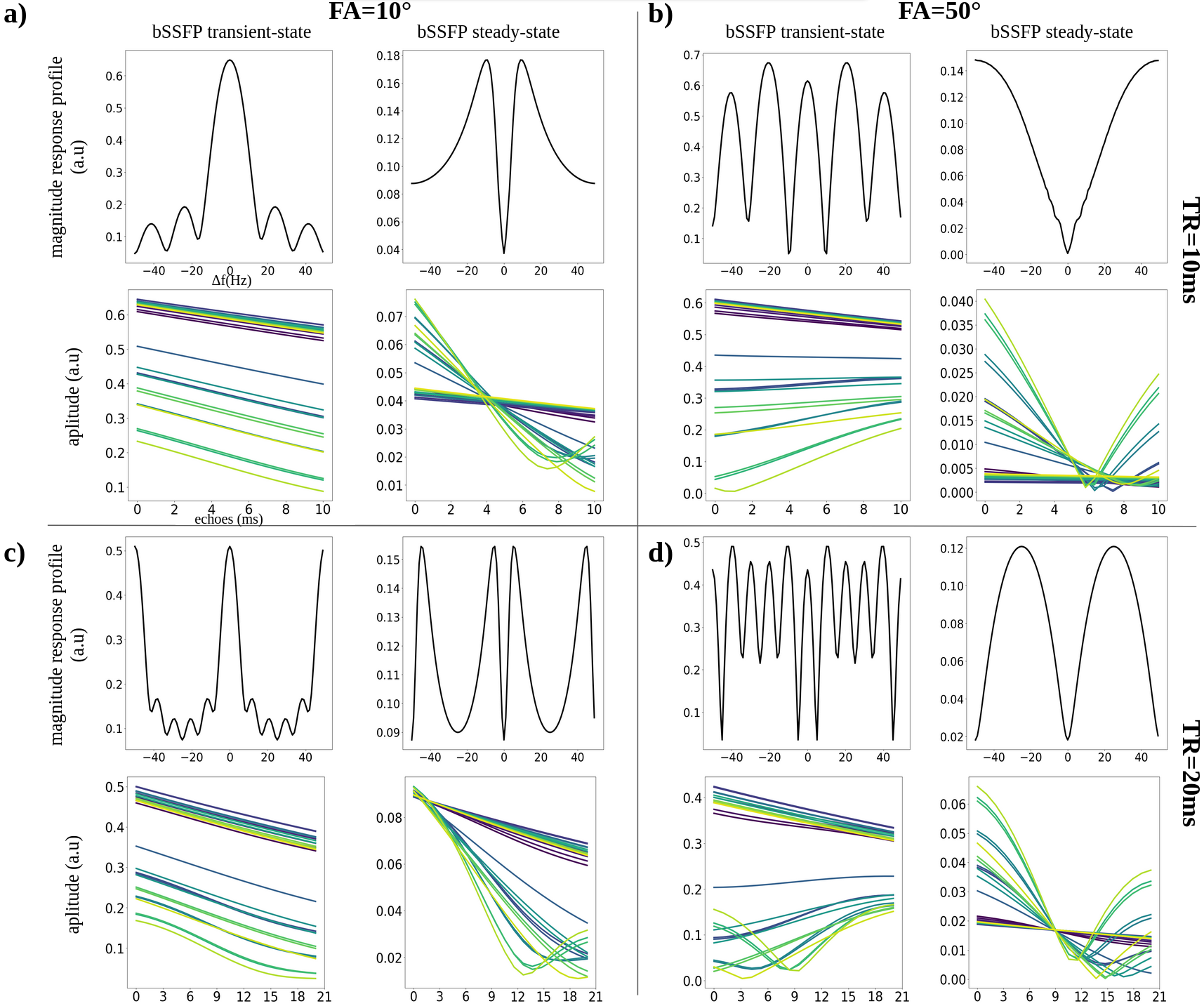}}
\caption{In silico study of the bSSFP response signal to the intra-voxel frequency distributions without contrast agent. Magnitude response profiles as well as sequence signal decays are shown for different sets of sequence parameters, for TE=TR/2, and average tissue properties of grey matter at 3T (GM, \tun=1300 ms, and \tdeux=80 ms) and for $\delta \! f$=0Hz. The transient state is represented by looking at echoes between two pulses at the very beginning of the sequence, while the steady state is assumed for echoes between two pulses after an important number of TRs depending on the sequence parameters.\label{fig5}}
\end{figure*}

In contrast, bSSFP signal profiles vary with sequence parameters. Figure \ref{fig5} shows that, even with fixed TR and FA, the profile differs between transient and steady states. For non-contrast distributions from \com{Figure S4}, signals may increase with echo time rather than decrease monotonically. Certain parameter combinations may produce spin echo-like signals at different echo times, and high flip angles can cause significant signal variability even at short echo times.

\begin{figure*}
\centerline{\includegraphics[  width=\textwidth  ]{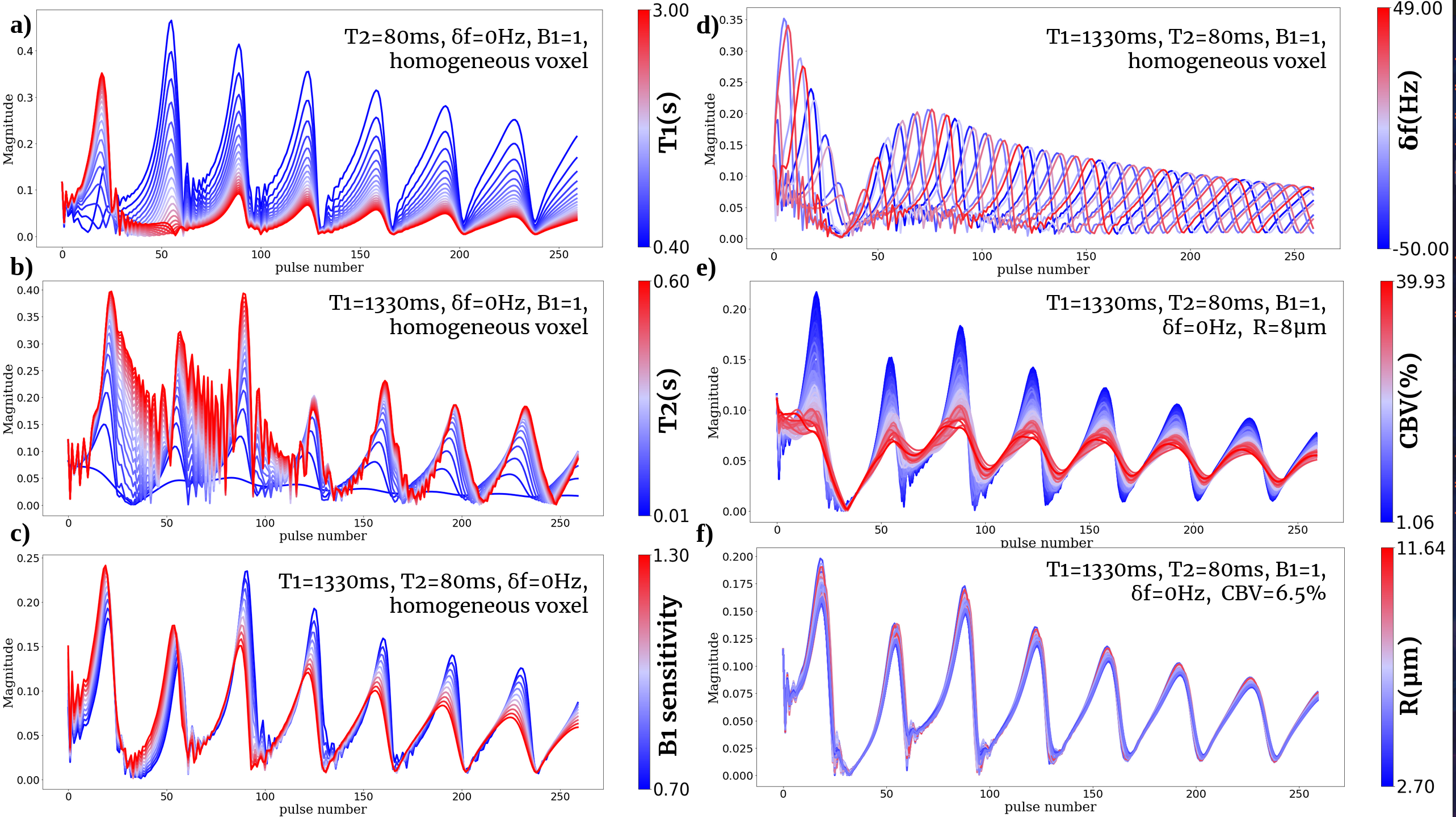}}
\caption{Sets of representative dictionary entries. For \tun, \tdeux, \bun, and $\delta \! f$ properties (a,b,c,d) the signal responses are plotted for a voxel devoid of the vasculature (single intra-voxel frequency, i.e. Dirac distribution), and only 1 of the 4 dictionary properties is varied whereas the remaining 3 are fixed. For CBV, and R properties (e,f), 1 of the 6 dictionary properties is varied whereas the remaining 5 are fixed.\label{fig6}}
\end{figure*}

The previous observations led to the design of the single echo, short TR, MRvF-bSSFP sequence that includes FA variations within transient and steady-state sections. This design enhances sensitivity to frequency distributions by allowing response profile variations over the signal acquisition time. 
Figure \ref{fig6} shows signal entries from subsections of the $Dico_{Micro}$ computed with the proposed sequence. Variations in signal evolution are evident with parameter changes, and different panels show limited cross-talk between parameters. Notably, changes in CBV cause distinct signal variations not captured by sensitivities to \tun, \tdeux, \bun{} or $\delta \! f$.

\subsection{In vivo acquisitions}

\begin{figure*}
\centerline{\includegraphics[  width=\textwidth  ]{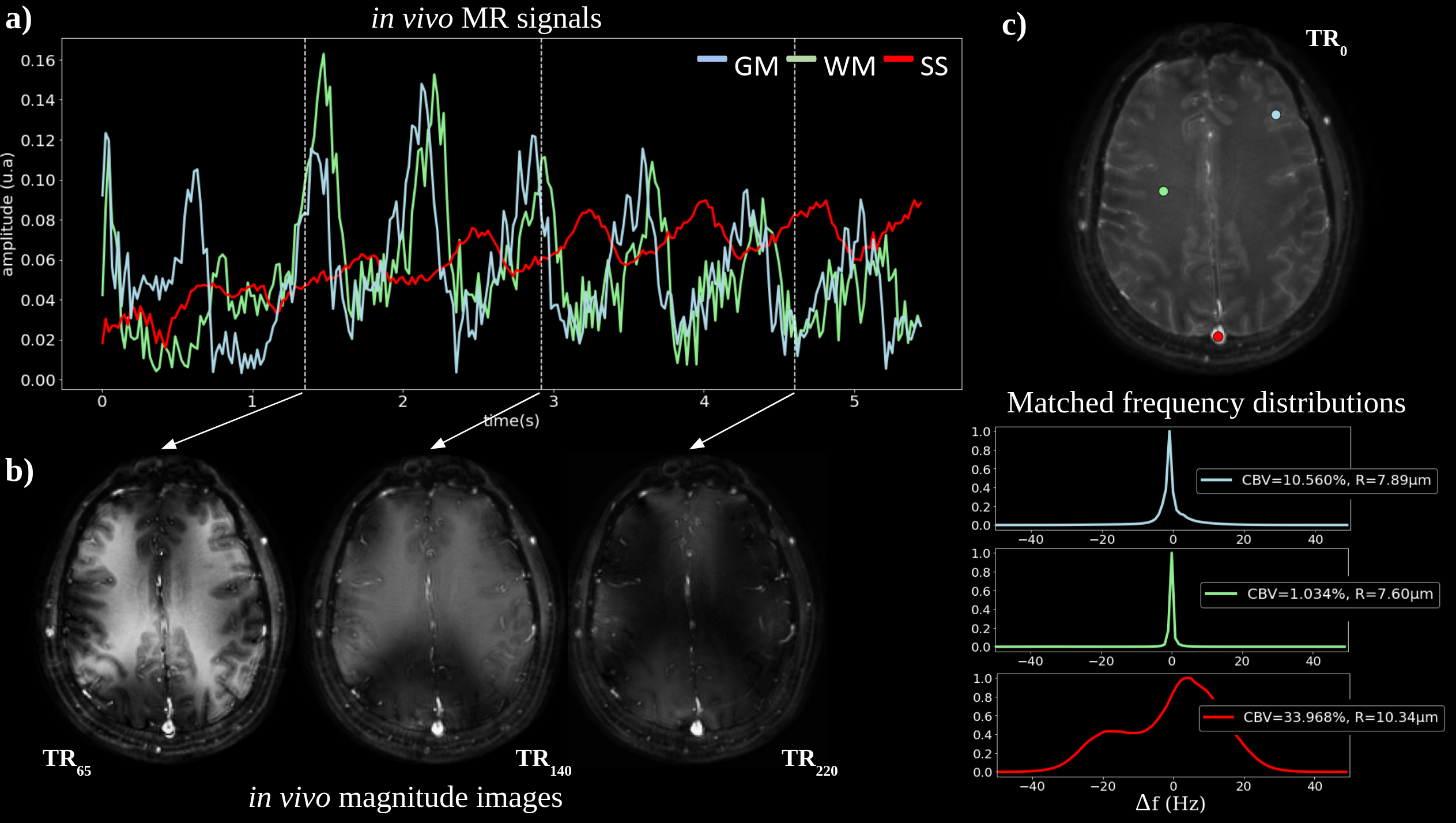}}
\caption{a) Example of in-vivo MR signals in grey matter (blue), white matter (green), and sagittal sinus (red) regions in one healthy volunteer. b) Magnitude images from in-vivo acquisition for different times of the sequence. c) Matched frequency distributions for the 3 tissue signals located by dots on the TR\textsubscript{0} magnitude image above.\label{fig7}}
\end{figure*}

Figure \ref{fig7} shows examples of in vivo MRF signals obtained in GM, WM, and SS regions in one healthy volunteer. Three magnitude images showing changes in contrast and bSSFP banding artifacts obtained at different pulse numbers are also shown. Figure \ref{fig7}c shows the magnitude image at first pulse TR$_0$ to localize the ROIs. The frequency distributions obtained after MRF matching and the associated (CBV, R) values are given. More estimated frequency distributions are shown in \com{Figure S5 (supp)}. Notably, the superior sagittal sinus vein shows a clearly different signal evolution as well as a larger matched frequency distribution with high estimated CBV and R values.

\begin{figure*}
\centerline{\includegraphics[  width=\textwidth  ]{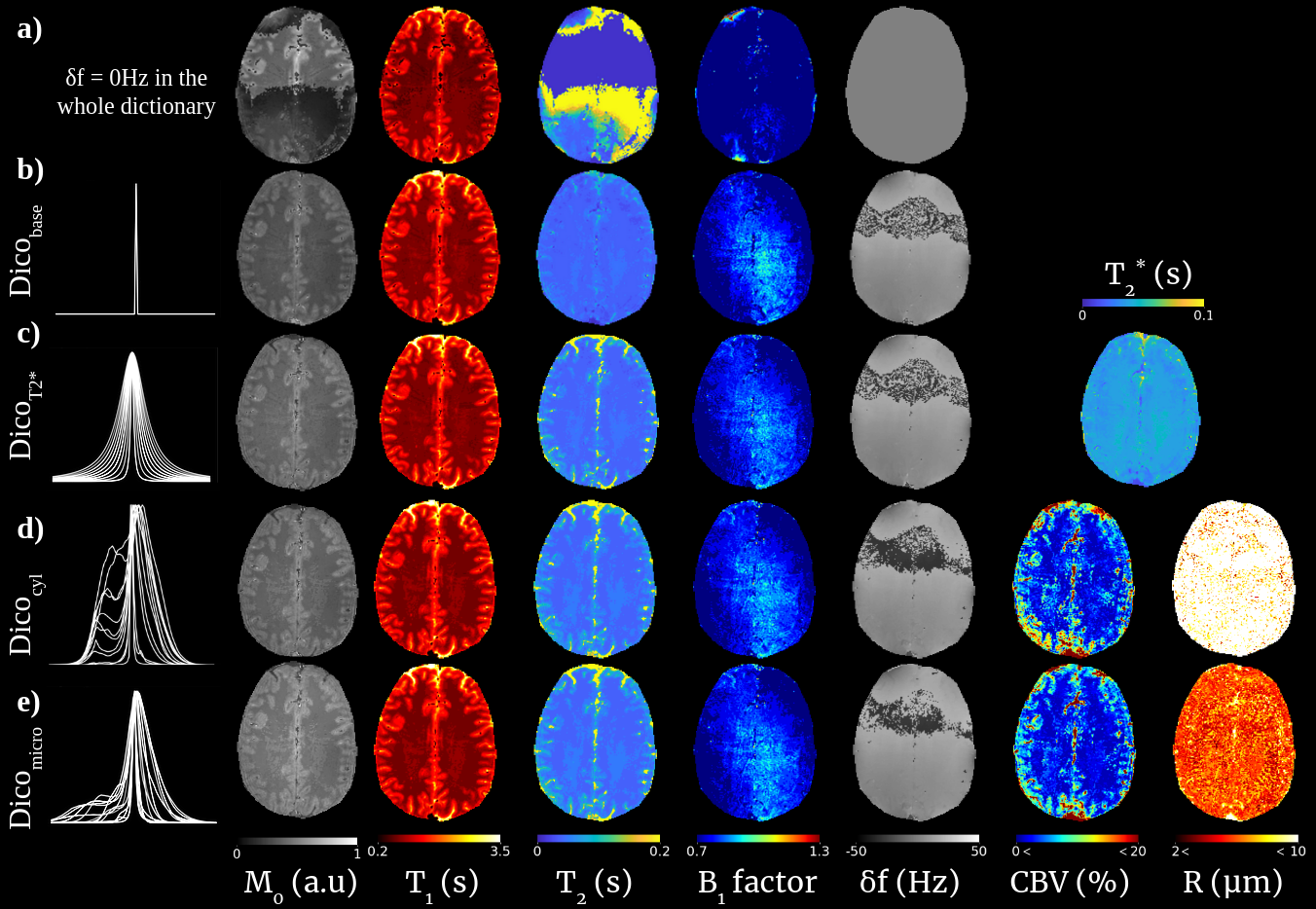}}
\caption{Matching results for our MRvF-bSSFP sequence with 5 different MRF dictionaries: a) defined on (\tun,\tdeux,\bun) parameters grid. b) defined on (\tun,\tdeux,\bun,$\delta \! f$) parameters grid. c) defined on (\tun,\tdeux,\bun,$\delta \! f$) parameter grid + Lorentzian distributions for \tdeuxetoile{} estimate. d) defined on (\tun,\tdeux,\bun,$\delta \! f$) parameters grid + frequency distributions from 3D cylindrical vascular model. e) defined on (\tun,\tdeux,\bun,$\delta \! f$) parameters grid + 3D frequency distributions from microscopy vascular model.\label{fig8}}
\end{figure*}

Figure \ref{fig8} shows results obtained with the MRvF-bSSFP sequence in another subject. Different dictionaries including increasingly realistic frequency distributions were used for the reconstructions. This highlights the sensitivity of the sequence to the frequency dimension, showing that \tun{} and \tdeux{} maps are not well estimated with unique $\delta \! f$ values. \tdeuxetoile{} maps and better \tdeux{} estimates are obtained when considering Lorentzian distributions. CBV and R values are obtained when considering vascular distributions with differences between the cylinder and realistic microscopy distributions. A clear contrast between CBV values in GM and WM can be observed. Larger CBV and R values are also found in large blood vessels such as the SS. Using realistic microscopy distributions, R values are lower than using cylinders and in the range of physiological values.

\begin{figure*}
\centerline{\includegraphics[  width=\textwidth  ]{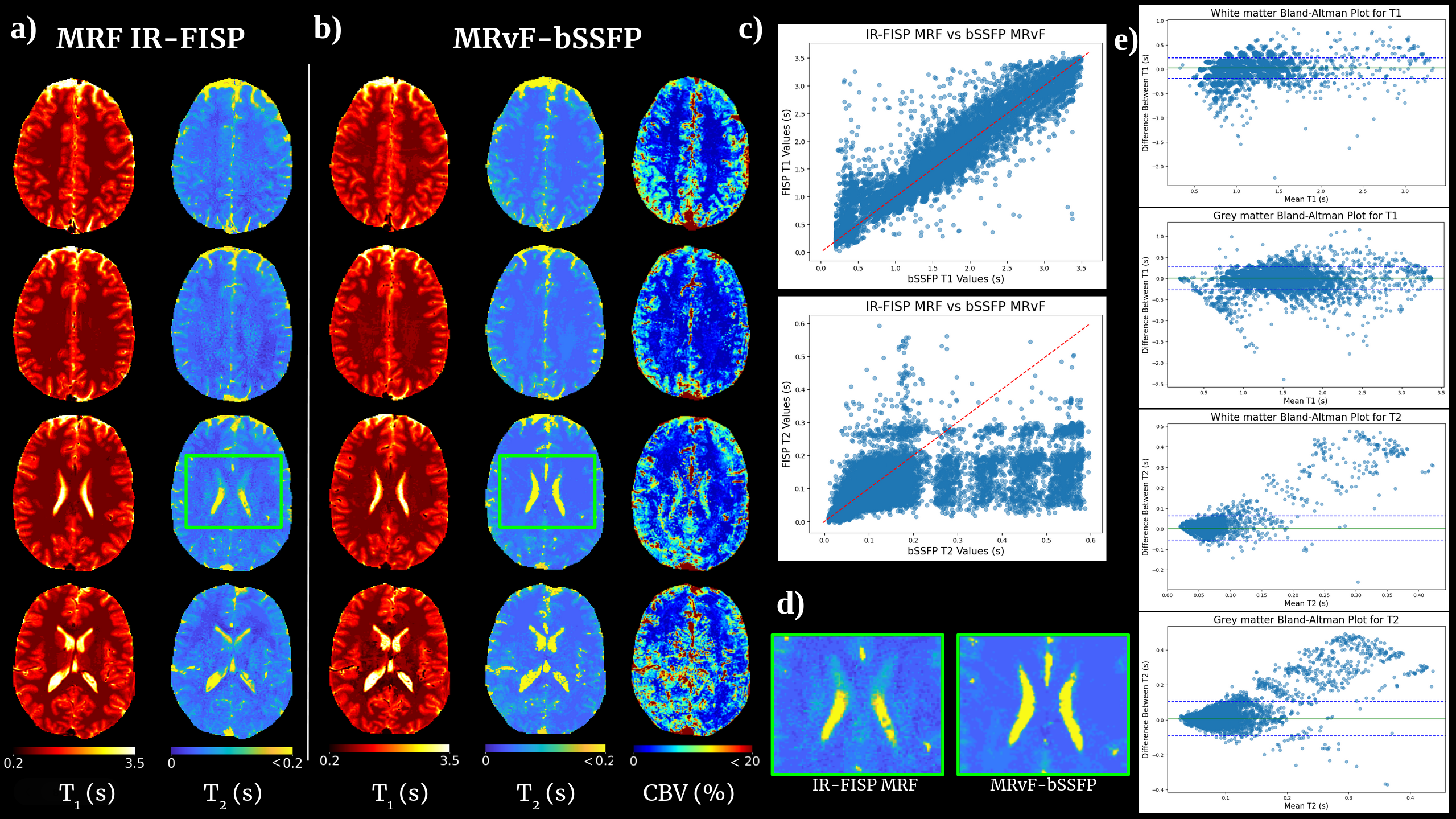}}
\caption{Comparison between (\tun,\tdeux) estimates from standard MRF IR-FISP sequence (a) and our approach based on MRvF-bSSFP (b) which also estimates (\bun,$\delta \! f$,CBV,R) using the $Dico_{Micro}$. c) in-vivo results comparing MRF IR-FISP and MRvF-bSSFP for all pixels \tun and \tdeux values.
d) Focus on the ventricles regions in the \tdeux estimates. e) Bland-Altman plots comparing MRF IR-FISP and MRvF-bSSFP. The dotted blue lines indicate the 95\% limits of agreement, and the solid green line indicates the mean bias.
\label{fig9}}
\end{figure*}

Relaxometry maps derived from the reference MRF IR-FISP sequence and our proposed MRvF-bSSFP sequence are compared in Figure \ref{fig9} a and b respectively, in which 4 slices of one subject are shown. Scatter plots and results from voxel-wise Bland Altman (BA) analyses are shown in (Figure \ref{fig9}c and e respectively). The scatter plot comparing \tun{} values shows a strong correlation with data points clustering around the line of identity (red dashed line). This indicates that \tun{} values from both sequences are generally in good agreement. Higher dispersion around the line of identity is seen in the upper limits of the \tdeux{} scatter plot which suggests differences in the quantitative maps for high \tdeux{} values as pointed out in Figure \ref{fig9}d. Better delineation of the ventricles regions is observed with the MRvF-bSSFP sequence. This is also shown by the noticeable spread around the bias in the BA plots for high \tdeux{} estimates in both WM and GM. CBV maps show degraded quality with possible shim and spiral artifacts when approaching deeper brain regions. Large vascular structures are however still visible.

The ROI analysis for the 6 volunteers is summarized in Table \ref{tab1}(mean $\pm$ standard error of the mean, N=6). Reference values obtained using the MRF IR-FISP sequence (\tun,\tdeux) and the MGRE sequence (\tdeuxetoile) are given for comparison. Ranges of values found in the literature are also given. The average measured GM/WM ratio for the CBV measurement over the six healthy volunteers is 1.91$\pm$0.31. Quantitative maps obtained for one slice of the six healthy volunteers are shown in \com{Figure S6 (supp)} of supplementary material.

\paragraph{High resolution maps and other contributions}\label{resultcontrib}

\begin{figure*}
\centerline{\includegraphics[  width=\textwidth  ]{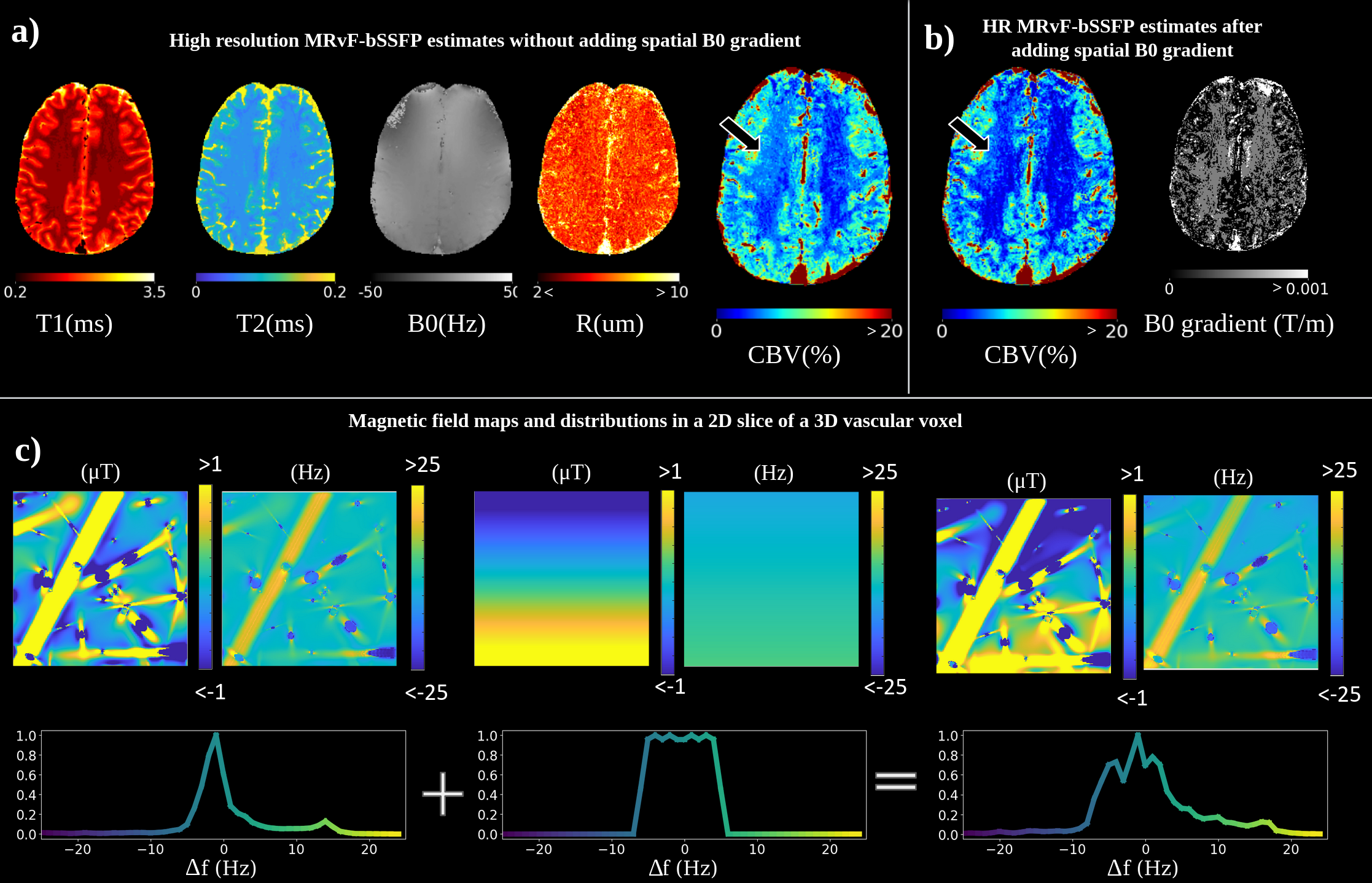}}
\caption{ a) MRvF-bSSFP estimates for the 2D cartesian acquisition using intra-voxel magnetic field distributions where only blood vessels contribute to field inhomogeneities (c). Convoluting top-hat function on frequency distributions accounts for spatial \bzero{} gradient in the X-direction. Adding function FWHM parameters in the dictionary allows the estimates of this spatial \bzero{} gradient strength as shown in b).
\label{fig10}}
\end{figure*}

In Figure \ref{fig10}a, we show the results obtained from the high-resolution cartesian MRF acquisition. The CBV map can be compared with and without taking into account an additional \bzero{} gradient dimension in the dictionary (see section \ref{matmet} and Figure \ref{fig10}b). Clear differences are observed in the white matter region and are highlighted by the black arrows. It is worth noting that the highest \bzero{} gradient values are mainly localized in the white matter region of the brain. Examples of magnetic fields' 2D \bzero{} spatial distributions and corresponding frequency distributions obtained in voxels containing vascular networks and one \bzero{} gradient are shown in Figure \ref{fig10}c. Note that in Figure \ref{fig10}b, only CBV and \bzero{} gradient estimates are shown, but the extended method allows the simultaneous estimate of \tun, \tdeux, \bun, $\delta \! f$, R, CBV, and \bzero{} gradient external contribution as well as \tdeuxetoile{} maps if using the Lorentzian dictionary.


\section{Discussion and Conclusions}
In this work, we proposed a new contrast-free method for the simultaneous quantification of relaxometry, magnetic fields, and microvascular parameters using the sensitivity of bSSFP sequences to sub-voxel frequency distributions in both transient and steady-state regimes. Because numerous parameters can influence these MR signals, we analyzed our results using the MRF framework, using dictionaries that include simulations on realistic microvascular networks. Our preliminary results obtained in 6 healthy subjects compare well with standard MRF FISP \tun{} and \tdeux{} estimates and are in the range of reported relaxometry literature values. Strong contrast between CBV estimates in WM and GM and a CBV ratio of 1.91 between these regions were expected\cite{cbv_bjornerud, cbv_knutsson, cbv_leenders, cbv_liu, cbv_uh, cbv_sakai}. Large values of CBV and R in voxels containing known large blood vessels are also encouraging. Results show good reproducibility of our findings across multiple subjects (\com{Figure S6 supp)}. 

However, \tdeuxetoile{} estimates using the MRvF-bSSFP methods are higher than the \tdeuxetoile{} values obtained using MGRE sequences, and quantitative values of CBV are on the high end of reported literature. Comparison of our technique against DSC in healthy tissues and lesions will be needed, although quantitative estimates of CBV using Gd injection might also be overestimated\cite{Wirestam2010}.  Pre-clinical studies in animal models of stroke or tumor could also be conducted, allowing comparison with steady-state perfusion measurements obtained with Ultra Small Particle of Iron Oxide (USPIO) CA as well as comparison with subsequent histological analysis\cite{valable_histo}. Changes in response to $CO_2$ gas challenges could also be used to induce temporal variations of CBV and oxygenation values.
 
Several optimizations are foreseeable. The model for MRvF simulations could be improved by adding additional magnetic field susceptibility sources. We already observed that simple additions of linear gradients in the dictionary had an impact on the vascular estimates. This could correspond to shim artifacts or the presence of myelin sheath in WM (see Figure S8 (supp)). Myelin fibers are known to contribute to the magnetic susceptibility of tissues and thus modify the underlying magnetic field distribution\cite{CHEN20131, Cottaar2021}. Intra-voxel frequency shifts can vary across different brain regions and have an impact on bSSFP frequency response profile asymmetries\cite{miller2010WM}. It could be possible to improve the simulations using higher-order shims and representations of microstructures in the virtual voxels. In the same way, nonheme iron in deep GM could be included. The vascular simulations could also be improved by considering realistic 3D networks from human brains and additional estimates such as vessel orientation and density could be considered. Finally, dictionary matching could be improved by taking into account the actual RF pulse profile in the simulations. It has been shown in \cite{ma2017slice, buonincontri2016mr} that including the slice profile effects in the dictionary simulations highly improves reconstructed maps quality for FISP-based sequences. This could enhance the measurement of relaxometry parameters as well as intra-voxel parameters. In \com{Figure S7 and Table S1 (supp)}, we show the impact of taking this slice profile effect into account before the convolution process. While interesting results can be observed on the \tdeuxetoile{} maps compared to the 2D multi-echo GRE reference, the effect is only partially corrected as the bSSFP slice profile also depends on off-resonance \cite{bssfp_sliceprofile2008}. Correctly incorporating RF profiles in our simulations would thus require a new type of efficient dictionary generation. 

Another way to improve the method is to work on data acquisition. In our proof of principle study, a 2D acquisition with high k-space sampling was chosen to avoid strong image artifacts. Long delays had to be put at the end of each spiral interleave to ensure the returns to equilibrium. A 3D version of the sequence with single shot acquisition would be much faster and would also avoid the need for above mentioned fastidious slice profile corrections. Yet, finding a better MRF acquisition pattern is not trivial given the high number of dimensions. Automatic procedures might be required \cite{lee2019flexible, COHEN201715, coudert_ismrm2022, jordan2021automated} and these algorithms could focus on optimized RF phase evolution that has already been shown to have a large impact in FISP sequences for diffusion or \tun{} estimates. The influence of TR or the addition of multiple echoes could also be explored to improve frequency distribution\cite{Schaper2022_purebssfp} or blood oxygenation sensitivities. However, one has to pay attention to the evolution of TR and FA which can also increase the sensitivity to water diffusion or magnetization transfer effects. In our first design, these contributions were minimized (\com{see Figure S1 (supp)}) using short TR, short TE, and low FA and allowed fast simulations that neglected the spatial arrangement of intra-voxel magnetic fields.

Given the potential new simulation models that would lead to larger dictionaries and longer MRF patterns for undersampled acquisition sequences, the MRF reconstruction process might have to switch from simple dot product matching to deep learning reconstruction with data compression\cite{Zhao2018}. Yet, it has been shown that subspace reconstruction of bSSFP sequences is not trivial\cite{MA_SVD}, and special DL network architectures might be needed to handle the nonlinear signal variations and large dimensions of the dictionaries\cite{barrier2024marvelmrfingerprintingadditional}.  Better reconstructions could also be obtained by using the complex data as input of the DL network\cite{Lu2022} as the phase information in bSSFP experiments has been shown to influence the frequency profile asymmetry of the sequence\cite{phasebSSFP}. 

It is important to note that, in this study, we are mainly focusing on deoxygenated blood, as the BOLD effect is primarily sensitive to changes in deoxyhemoglobin. By focusing on deoxyhemoglobin, our results rely on venous CBV, excluding arterial contributions. 

Once properly validated, our method could help clinical investigations of several cerebrovascular pathologies including stroke, or cancer.


\subsection*{Author contributions}
All authors listed have made substantial, direct, and intellectual contributions, proofread and corrected the final manuscript, and approved it for publication. TCo, TCh, and EB took part in the conception of this study. TCo, AD, TCh, AB, and JW took part in the conception and realization of the numerical simulations and data analysis codes. LLa, MD, and BL assist with the review and editing of the manuscript. LLe share expertise and guidance on the medical aspects and the study's region of interest. MD and LLa provide support and assistance on the MRI software machine. TCo and TCh wrote the manuscript.

\subsection*{Financial disclosure}

The MRI facility IRMaGe is partly funded by the French program “Investissement d’avenir” run by the French National Research Agency, grant “Infrastructure d’avenir en Biologie et Santé”. [ANR-11-INBS-0006]
The project is supported by the French National Research Agency. [ANR-20-CE19-0030 MRFUSE]

\subsection*{Conflict of interest}
Mariya Doneva is an employee of Philips GmbH Innovative Technologies.
\newpage
\vfill\pagebreak

\section*{Supporting information}
The following supporting information is available as part of the online article:

\vskip\baselineskip\noindent
\textbf{Figure S1.} shows the effect of water diffusion on the signal shape simulated using the MRvF-bSSP sequence.

\noindent
\textbf{Figure S2.} shows simulation parameters for the three dictionaries presented in the Methods section.

\noindent
\textbf{Figure S3.} shows intra-voxel frequency distributions for voxels of varying sizes synthetically generated with cylindrical vessels.

\noindent
\textbf{Figure S4.} shows intra-voxel frequency distributions without (a) and with (b) the presence of a USPIO contrast agent.

\noindent
\textbf{Figure S5.} shows in vivo matched results of intra-voxel frequency distributions. 

\noindent
\textbf{Figure S6.} shows in vivo quantitative maps for the 6 healthy volunteers of our study.

\noindent
\textbf{Figure S7.} shows in vivo quantitative maps for one subject using Lorentzian distributions for \tdeuxetoile{} quantification with and without slice profile effect. 

\noindent
\textbf{Table S1.} shows ROIs values for \tdeuxetoile{} MRF-bSSFP estimation with and without slice profile effect simulation.

\noindent
\textbf{Figure S8.} propose a comparison of the obtained "\bzero{} gradient" with a PV-MRF computed Myelin Water Fraction map.

\vspace*{6pt}

\newpage

\listoffigures
\listoftables

\begin{table*}[t] 
\caption{Region of interest analysis for in vivo \tun, \tdeux, \tdeuxetoile, CBV and R values.\label{tab1}}
\begin{tabular*}{\textwidth}{@{\extracolsep\fill}l*{4}{c}@{\extracolsep\fill}}
\toprule
\textbf{Parameter} &
  \textbf{Tissue} &
  \textbf{MRvF-bSSFP} &
  \textbf{References$^{1,2}$} &
    \textbf{Litterature} \\ \hline
\tun(ms) &
  {\begin{tabular}[c]{@{}c@{}}WM\\ GM\end{tabular}} &
  \begin{tabular}[c]{@{}c@{}}943 $\pm$ 63\\ 1334 $\pm$ 37\end{tabular} &
  \begin{tabular}[c]{@{}c@{}}919 $\pm$ 44\\ 1341 $\pm$ 46\end{tabular} &
   \begin{tabular}[c]{@{}c@{}}$\sim790-1080$ \\ $\sim1180-1820$\citep{wansapura1999nmr,stanisz2005t1, lu2005routine, jiang2015mr, Ma2013} \end{tabular}
   \\ \hline
\tdeux(ms) &
  {\begin{tabular}[c]{@{}c@{}}WM\\ GM\end{tabular}} &
  \begin{tabular}[c]{@{}c@{}}47 $\pm$ 5\\ 68 $\pm$ 7\end{tabular} &
  \begin{tabular}[c]{@{}c@{}}46 $\pm$ 3\\ 64 $\pm$ 4\end{tabular} &
   \begin{tabular}[c]{@{}c@{}}$\sim56-84$ \\ $\sim70-130$\citep{wansapura1999nmr,stanisz2005t1, lu2005routine,jiang2015mr,Ma2013}  \end{tabular}
   \\ \hline
\tdeuxetoile(ms) &
  {\begin{tabular}[c]{@{}c@{}}WM\\ GM\end{tabular}} &
  \begin{tabular}[c]{@{}c@{}}37 $\pm$ 2\\ 37 $\pm$ 2\end{tabular} &
  \begin{tabular}[c]{@{}c@{}}49 $\pm$ 2\\ 51 $\pm$ 2\end{tabular} &
   \begin{tabular}[c]{@{}c@{}}$\sim45-48$ \\ $\sim42-52$ \citep{ni2015comparison,Peters2007}  \end{tabular}
   \\ \hline
CBV(\%) &
  {\begin{tabular}[c]{@{}c@{}}WM\\ GM\\ SS\end{tabular}} &
  \begin{tabular}[c]{@{}c@{}}4.72 $\pm$ 0.87\\ 8.65 $\pm$ 1.38\\ 36.70 $\pm$ 2.87 \end{tabular} & \begin{tabular}[c]{@{}c@{}} --- \\ --- \\ ---\end{tabular} 
  & \begin{tabular}[c]{@{}c@{}}$\sim1.7-3.6$ \\ $\sim3.0-8.0$ \citep{cbv_bjornerud,cbv_knutsson,cbv_leenders,cbv_liu,cbv_uh,cbv_sakai}\\ --- \end{tabular}
   \\ \hline
R($\mu$m) &
  {\begin{tabular}[c]{@{}c@{}}WM\\ GM\\ SS\end{tabular}} &
  \begin{tabular}[c]{@{}c@{}}5.85 $\pm$ 0.07\\ 6.27$\pm$ 0.21\\ 10.20 $\pm$ 0.44 \end{tabular} &
  \begin{tabular}[c]{@{}c@{}} --- \\ --- \\ ---\end{tabular} 
  & \begin{tabular}[c]{@{}c@{}}6.8 $\pm$ 0.3 \\ 7.3 $\pm$ 0.3 \citep{Delphin2023} \\ ---\end{tabular}
   \\ \hline
\bottomrule
\end{tabular*}

\end{table*}

\FloatBarrier
\bibliographystyle{plain}
\bibliography{biblio}

\includepdf[pages=-]{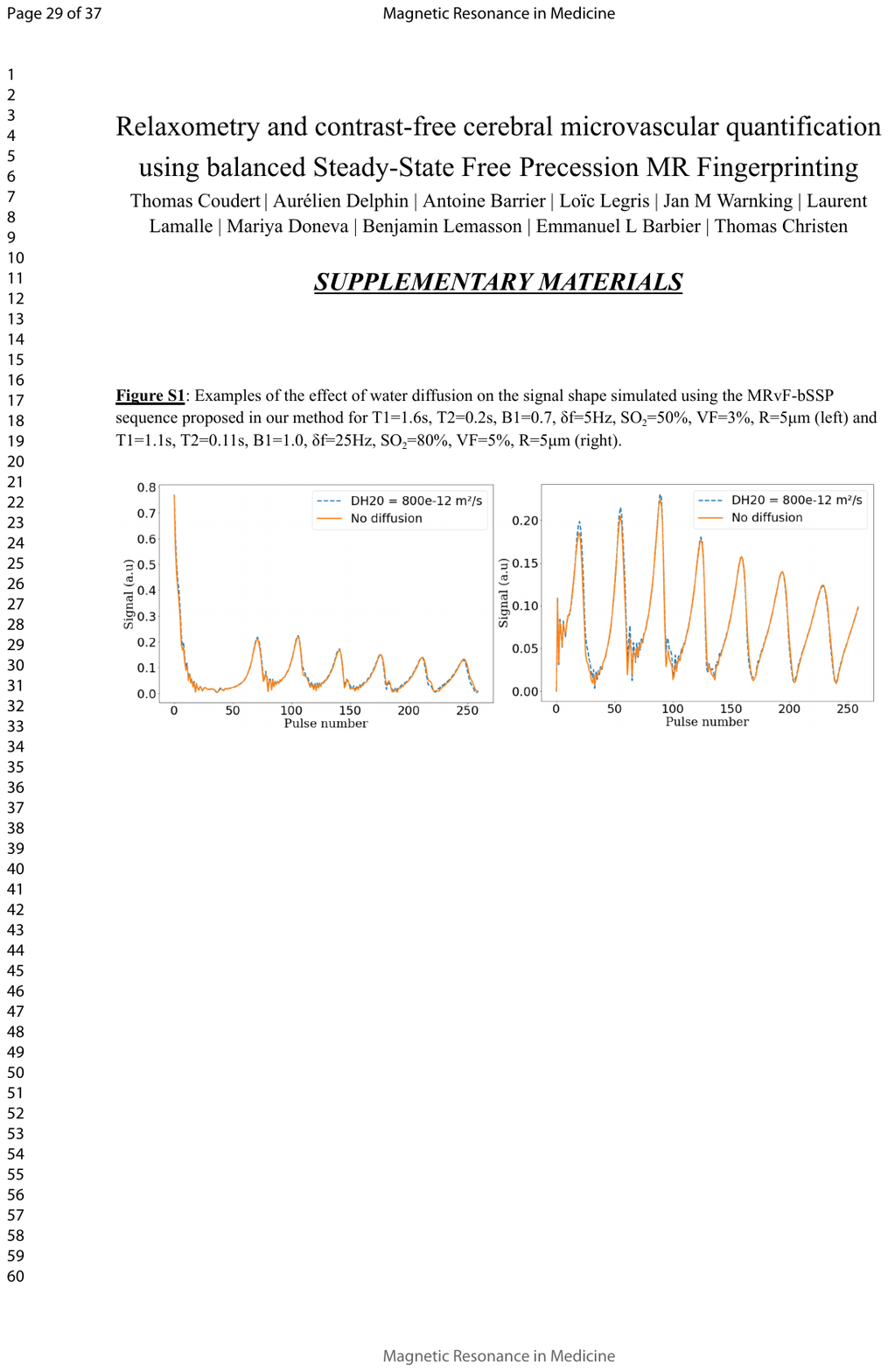} %
\end{document}